\documentclass[11pt]{article}
\usepackage{moriond}

\def\be{\begin{equation}}
\def\ee{\end{equation}}
\def\bea{\begin{eqnarray}}
\def\eea{\end{eqnarray}}

\newcommand{\Dfb}{\mbox{$\raisebox{2mm}{\boldmath ${}^\leftrightarrow$}
\hspace{-4mm} D$}}
\newcommand{\Dfba}{\mbox{$\raisebox{2mm}{\boldmath ${}^\leftrightarrow$}\hspace{-4mm} D^a$}}

\newcommand{\Photo}{}

\begin{document}
\vspace*{4cm}
\title{Robust determination of the scalar boson couplings
\footnote{Talk given by O. \'Eboli at the Rencontre de Moriond EW 2013.}}

\author{Tyler Corbett$^\dagger$, 
        O.\ J.\ P.\ \'Eboli$^\clubsuit$,
        J.\ Gonzalez--Fraile$^\diamondsuit$, and 
        M.\ C.\ Gonzalez--Garcia$^{\dagger,\diamondsuit,\spadesuit}$\\ }

\address{%
  $^\dagger$C.N.~Yang Institute for Theoretical Physics, SUNY,
  Stony Brook, NY 11794-3840, USA. \\
  $^\clubsuit$Instituto de F\'{\i}sica, Universidade de S\~ao Paulo, S\~ao
  Paulo -- SP, Brazil. \\
  $^\diamondsuit$Departament d'Estructura i Constituents de la Mat\`eria and
  ICC-UB,
Barcelona,
  Spain.\\
  $^\spadesuit$Instituci\'o Catalana de Recerca i Estudis Avan\c{c}ats (ICREA).
}

\maketitle\abstracts{
We make a study of the indirect effects of new physics on the phenomenology of
the ``Higgs-like'' particle.  Assuming that the
recently observed state is a light electroweak doublet scalar and
the $SU(2)_L \otimes U(1)_Y$ symmetry is realized linearly, we
parametrize these effects in a model independent way in terms of an effective Lagrangian at the
electroweak scale.
We choose the dimension--six operator that allows us to
better use all the available data to constrain the coefficients of
the dimension-six operators.
Subsequently we perform a global 6--parameter fit which allows for simultaneous
determination of the standard model scalar couplings to electroweak gauge bosons, gluons,
bottom quarks, and tau leptons. The results
are based on the data released at Moriond 2013.
Moreover, our formalism leads to strong constraints on the electroweak
triple gauge boson couplings. }

\section{Low energy effective Lagrangian: the right of choice}

The 2012 discovery of the state resembling the standard model scalar
(SMS) marks the beginning of the direct study of electroweak
symmetry breaking (EWSB)~\cite{descoberta}. To
determine whether this new particle is the state predicted by the
standard model (SM) we still must determine its properties like spin,
parity, and couplings. Here we use a bottom--up approach to study the
SMS couplings by parametrizing the deviations of the SM predictions by
an effective Lagrangian.

We assume that the newly observed state is part of an electroweak
scalar doublet and the gauge $SU(2)_L \otimes U(1)_Y$ is linearly
realized. The lowest order operators that modify the SMS couplings are
of dimension--six~\cite{dim6}. It is well known~\cite{59} that there
are 59 independent dimension--six operators up to flavor and Hermitian
conjugation.  However, there is
a freedom in the choice of the basis of operators since operators
connected by the equations of motion lead to the same $S$--matrix
elements~\cite{eom}.  Thus, the determination of physical observables
like branching ratios or production cross sections would be
independent of the basis  choice. Nevertheless independent does not
mean equivalent. As a result of this reasoning, we propose that in the
absence of theoretical prejudices it turns is beneficial to use
a basis allowing for the use of largest dataset in our analyses.
Therefore, a sensible (and certainly technically convenient) choice is
to leave in the basis  those
operators which are directly related to the existing data, for example
the bulk of precision electroweak measurements which
have with the establishment of the SM.

In our analysis we characterize deviations from the predictions of the SM as
\begin{equation}
{\cal L}_{\rm eff} = \sum_n \frac{f_n}{\Lambda^2} {\cal O}_n \;\; ,
\label{l:eff}
\end{equation}
where ${\cal O}_n$ are the dimension--six operators which involve gauge--bosons,
and/or fermionic fields, and the SMS, with couplings $f_n$ and where
$\Lambda$ is the characteristic scale. We assume ${\cal O}_n$ operators to be $P$ and $C$ even and the
conservation of baryon and lepton numbers. Our fit to the available datasets
leads to constraints on $f_n/\Lambda^2$ that can be easily translated
into SMS properties and into bounds on triple gauge boson couplings.
The basic building blocks for the dimension--six operators are the SMS
doublet $\Phi$ and its covariant derivative, $D_\mu\Phi=
\left(\partial_\mu+i \frac{1}{2} g' B_\mu + i g \frac{\sigma_a}{2}
W^a_\mu \right)\Phi $ in our conventions, as well as, the hatted field
strengths defined as $\hat{B}_{\mu \nu} = i \frac{g'}{2} B_{\mu \nu}$
and $\hat{W}_{\mu\nu} = i \frac{g}{2} \sigma^a W^a_{\mu\nu}$. The gauge couplings of
$SU(2)_L$ ($U(1)_Y$) are denoted as $g$ ($g^\prime$) and
the Pauli matrices by $\sigma^a$. The fermionic degrees of freedom are
the lepton doublets $L$, the quark doublets $Q$ and the $SU(2)_L$
singlet fermions $f_R$.

Here we directly present our choice of basis, however the detailed
discussion of this choice can be found in this
reference~\cite{ourWork}. In this basis the bosonic operators
modifying the SMS interactions with the gauge bosons are
\begin{equation}
\begin{array}{lll}
 {\cal O}_{GG} = \Phi^\dagger \Phi \; G_{\mu\nu}^a G^{a\mu\nu}  \;\;,
& {\cal O}_{WW} = \Phi^{\dagger} \hat{W}_{\mu \nu} 
 \hat{W}^{\mu \nu} \Phi  \;\; , 
&  {\cal O}_{BW} =  \Phi^{\dagger} \hat{B}_{\mu \nu} 
 \hat{W}^{\mu \nu} \Phi \;\; ,
\\
& &
\\
{\cal O}_W  = (D_{\mu} \Phi)^{\dagger} 
  \hat{W}^{\mu \nu}  (D_{\nu} \Phi) \;\; ,

& {\cal O}_B  =  (D_{\mu} \Phi)^{\dagger} 
  \hat{B}^{\mu \nu}  (D_{\nu} \Phi)  \;\; ,
& {\cal O}_{\Phi,1} 
=  \left ( D_\mu \Phi \right)^\dagger \Phi\  \Phi^\dagger
\left ( D^\mu \Phi \right ) \;\; , 
\end{array}
\label{eff:bosons}  
\end{equation}
while the dimension--six operators relevant for the SMS interactions
with fermions are
\begin{equation}
\begin{array}{l@{\hspace{1cm}}l@{\hspace{1cm}}l}
{\cal O}_{e\Phi,ij}=(\Phi^\dagger\Phi)(\bar L_{i} \Phi e_{R_j}) ,
& 
{\cal O}^{(1)}_{\Phi L,ij}=\Phi^\dagger (i\, \Dfb_\mu \Phi) 
(\bar L_{i}\gamma^\mu L_{j}) ,
& 
{\cal O}^{(3)}_{\Phi L,ij}=\Phi^\dagger (i\,{\Dfba}_{\!\!\mu} \Phi) 
(\bar L_{i}\gamma^\mu \sigma_a L_{j}) , \\
{\cal O}_{u\Phi,ij}=(\Phi^\dagger\Phi)(\bar Q_{i} \tilde \Phi u_{R_j}) ,
& 
{\cal O}^{(1)}_{\Phi Q,ij}=\Phi^\dagger (i\,\Dfb_\mu \Phi)  
(\bar Q_i\gamma^\mu Q_{j}) ,
& 
{\cal O}^{(3)}_{\Phi Q,ij}=\Phi^\dagger (i\,{\Dfba}_{\!\!\mu} \Phi) 
(\bar Q_i\gamma^\mu \sigma_a Q_j) ,\\
{\cal O}_{d\Phi,ij}=(\Phi^\dagger\Phi)(\bar Q_{i} \Phi d_{Rj}) ,
& 
{\cal O}^{(1)}_{\Phi e,ij}=\Phi^\dagger (i\Dfb_\mu \Phi) 
(\bar e_{R_i}\gamma^\mu e_{R_j})  ,
& 
\\
& {\cal O}^{(1)}_{\Phi u,ij}=\Phi^\dagger (i\,\Dfb_\mu \Phi) 
(\bar u_{R_i}\gamma^\mu u_{R_j}) ,
& \\

& {\cal O}^{(1)}_{\Phi d,ij}=\Phi^\dagger (i\,\Dfb_\mu \Phi) 
(\bar d_{R_i}\gamma^\mu d_{R_j}) ,
& \\

& {\cal O}^{(1)}_{\Phi ud,ij}=\tilde\Phi^\dagger (i\,\Dfb_\mu \Phi) 
(\bar u_{R_i}\gamma^\mu d_{R_j}) ,
& 
\end{array}
\label{eff:fermions}
\end{equation}
where we define $\tilde \Phi=\sigma_2\Phi^*$,
$\Phi^\dagger\Dfba_{\!\!\mu}\Phi= \Phi^\dagger \sigma^a
D_\mu \Phi-(D_\mu\Phi)^\dagger\sigma^a \Phi$, and  $\Phi^\dagger\Dfb_\mu\Phi= \Phi^\dagger D_\mu\Phi-(D_\mu\Phi)^\dagger
\Phi$and where we denote the
family indices by $i, j$. In addition to the above operators the
dimension--six basis also contains four--fermion interactions, dipole
operators, as well as, the operator ${\cal O}_ {WWW}$ that leads to
anomalous triple gauge couplings but does not modify the SMS
interactions~\cite{ourWork}.

One important property of this choice of basis is the presence of the
operators ${\cal O} _W$ and ${\cal O}_B$ that modify the SMS couplings
to gauge boson pairs, as well as, the triple electroweak gauge
couplings (TGC).  This allows us to use the available TGC data to
constrain the SMS properties.  Moreover, the SMS data also have an
impact on the present determination of the TGC as shown
below~\cite{ourWork,SMS-TGC}.

Now we take advantage of and apply all available experimental information to
the effect of reducing the number of relevant parameters in the analysis of
the SMS data:
\begin{itemize}

\item Considering that the $Z$ couplings to fermions are in agreement
  with the SM at the level of per mil~\cite{ALEPH}, the operators modifying these couplings will have coefficients
  so constrained that
  they will have no impact in the SMS physics with the present
  statistics. Consequently, we remove the operators $( {\cal
    O}^{(1)}_{\Phi f}, {\cal O}^{(3)}_{\Phi f} )$ from our analyses.

\item The precision electroweak parameters $S$ and $T$ strongly
  constrain the coefficients of ${\cal O}_{BW}$ and ${\cal
    O}_{\Phi,1}$, therefore, we also neglect their contributions.

\item The off--diagonal part of ${\cal O}_{f \Phi}$ is strongly
  constrained by data on low--energy flavor--changing interactions.
  Consequently we also discard them from our basis.

\item Flavor diagonal ${\cal O}_{f \Phi}$ from the first and second
  generations only have an effect on the present Higgs data via their contribution
  to the SMS--gluon--gluon and SMS--$\gamma$--$\gamma$ vertex at the one
  loop level.  The form factors are very suppressed for light fermion loops
  and correspondingly their effect is totally negligible in the
  analysis.  Therefore, we keep the fermionic operators ${\cal
    O}_{e \Phi, 33}$, ${\cal O}_{u \Phi, 33}$ and ${\cal O}_{d \Phi,
    33}$ only.

\item Tree level information concerning $h t \bar{t}$ production has
  very large errors still. So the parameter $f_{u\Phi,33}$ effectively
  contributes only to the one--loop SMS couplings gluon and photon pairs.  
  Presently these contributions can be absorbed into the
  redefinition of the parameters $f_{WW}$ and $f_g$ , therefore, we take
  $f_{u\Phi,33}\equiv f_{\rm top} \equiv 0$. In the future, when a larger luminosity is
  accumulated, it will be necessary to reintroduce $f_{\rm top}$ as one
  of the parameters in the fit.

\end{itemize}

Therefore, at the end of the day, the effective Lagrangian relevant to
our analyses is
\begin{equation}
{\cal L}_{eff} = - \frac{\alpha_s v}{8 \pi} \frac{f_g}{\Lambda^2} 
{\cal O}_{GG}
+ \frac{f_{WW}}{\Lambda^2} {\cal O}_{WW}
+ \frac{f_{W}}{\Lambda^2} {\cal O}_{W}
+ \frac{f_{B}}{\Lambda^2} {\cal O}_{B}
+ \frac{f_{\rm bot}}{\Lambda^2} {\cal O}_{d\Phi,33}
+ \frac{f_{\tau}}{\Lambda^2} {\cal O}_{e\Phi,33}
\;\; ,
\label{ourleff}
\end{equation}
that contains 6 unknown parameters $(f_g, f_{WW}, f_W, f_B, f_{\rm
  bot}, f_\tau)$.

To obtain the present constraints on the six unknown
parameters we construct a chi-square function~\cite{ourWork} using all
available data on the SMS production and decay coming from LHC and
Tevatron~\cite{SMSdata} and also on TGC~\cite{TGCdata} and electroweak
precision data (EWPD).  The details of the statistical analyses are
presented in reference~\cite{ourWork}.

\section{Results}

\begin{figure} [h!]
\centerline{\includegraphics[width=0.9\linewidth]{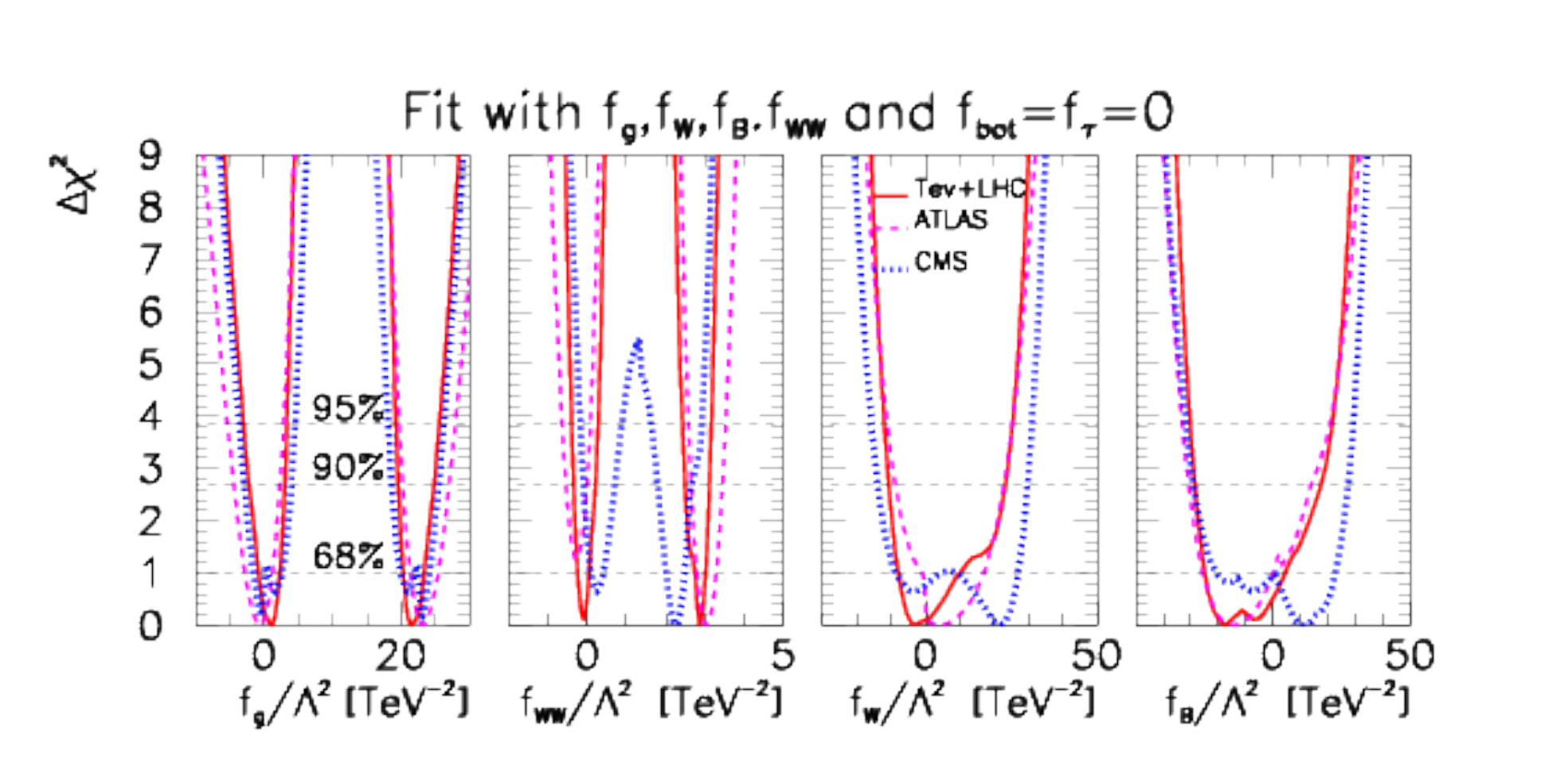}}
 \caption{$\Delta \chi^2 $ as a function of $f_{g}$, $f_{WW}$,
   $f_{W}$, and $f_B$ assuming $f_{\rm bot} = f_\tau = 0$. These
   panels contain three lines: the dashed (dotted) line are obtained
   using only the ATLAS (CMS) data and the solid line stands for the
   result using all available SMS data from Tevatron, ATLAS, and CMS. }
\label{fig:1}
\end{figure}

In order to compare the bounds on the SMS couplings coming from ATLAS
and CMS data we considered a scenario where the SMS couplings to
fermions take on their SM values, {\em i.e.}, we set $f_{\rm bot} =
f_\tau = 0$ and we fit the available data with $\{ f_g, f_W, f_B, f_{WW}
\}$ as the relevant free independent parameters.  Figure~\ref{fig:1} depicts the
$\Delta\chi^2$ as a function of each of the four free parameters after
having marginalized over the three unshown parameters.  As we can see in the
left panel, $\Delta\chi^2$ as a function of $f_g$ possesses two
degenerate minima caused by interference between SM and the anomalous
contributions. For the case of the chi--square dependence on $f_{WW}$
again there is an interference between SM and anomalous contributions,
however, the degeneracy of the minima is lifted as a result of the $f_{WW}$
coupling contributing to SMS decays into photons, $WW^*$ and $ZZ^*$, as
well as in $Vh$ associated and vector boson fusion production
mechanisms. Moreover, we can see from this figure that the CMS, ATLAS
and combined data exhibit a similar chi--square behavior with respect
to the fitting parameters and that the ATLAS and CMS data lead to
similar bounds on the SMS couplings at 90\% CL.

\begin{figure}
\centering
\centerline{\includegraphics[width=\linewidth]{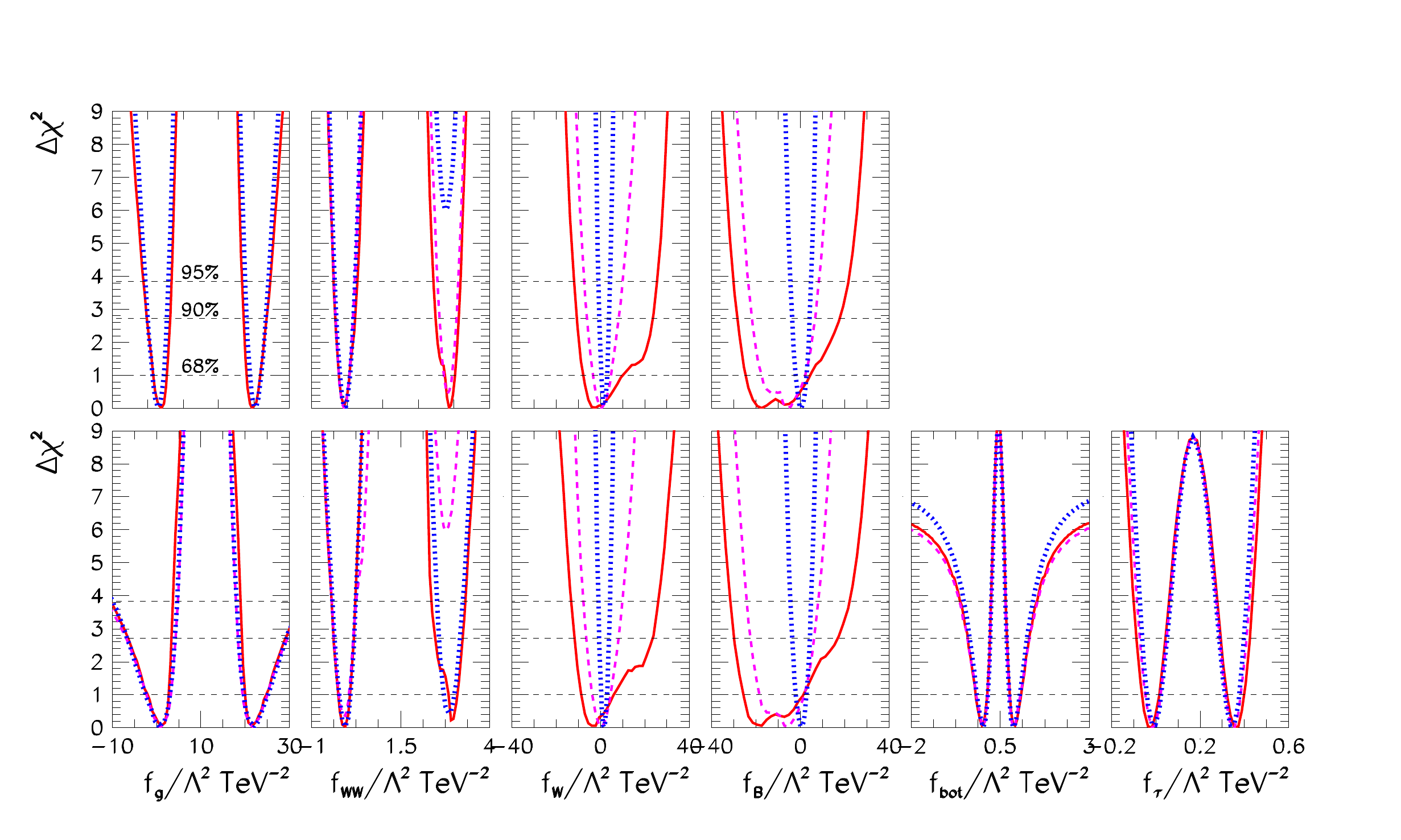}}
\label{fig:2}
 \caption{$\Delta\chi^2$ as a function of the fit parameters when
   all SMS collider (ATLAS, CMS and Tevatron) data (solid red
   line) are considered, SMS collider and TGC data (dashed purple line) and SMS
   collider, TGC and electroweak precision data (dotted blue line).
   The columns display the $\Delta\chi^2$ as a function of the fit
   parameter shown at the bottom of the column.  In the first row we
   use $f_g$, $f_{WW}$, $f_W$, and $f_B$ as fitting parameters with
   $f_{\rm bot} = f_\tau =0$, while in panels of the lower row we fit
   the data in terms of $f_g$, $f_{WW}$, $f_W$, $f_B$, $f_{\rm bot}$,
   and $f_\tau$. }
\end{figure}

The effect of combining the SMS data with the TGC data and EWPD is
presented in Fig.~\ref{fig:2} where a different set of free parameters
was used for each row. In the upper row of the figure the SMS
couplings to fermions are the SM ones while in the lower row the set
of fitting parameters is augmented with the inclusion of anomalous
bottom and tau couplings, $f_{\rm bot}$ and $f_\tau$, {\em i.e.} we
perform a six--parameter fit in $\{ f_g, f_W, f_B, f_{WW}, f_{\rm
  bot}, f_\tau \}$.  When including EWPD a scale of 10 TeV was assumed
in the evaluation of logarithms appearing in the expressions for
$S$, $T$ and $U$; see reference~\cite{ourWork} for further details.
Comparing the panels in the same column we can see that the impact of
the different datasets is similar in the two scenarios depicted in
this figure.  Because $f_B$ and $f_W$ are the only fit parameters
modifying the TGC at tree level, they show the largest
impact of the TGC data, particularlly $f_W$.  Moreover, the inclusion of
the EWPD in the fit has the affect of reducing the errors on $f_B$ and
$f_W$ significantly, in addition to lifting the near degeneracy on $f_{WW}$.

\begin{figure}[ht!]
  \centering
  \includegraphics[width=0.65\textwidth]{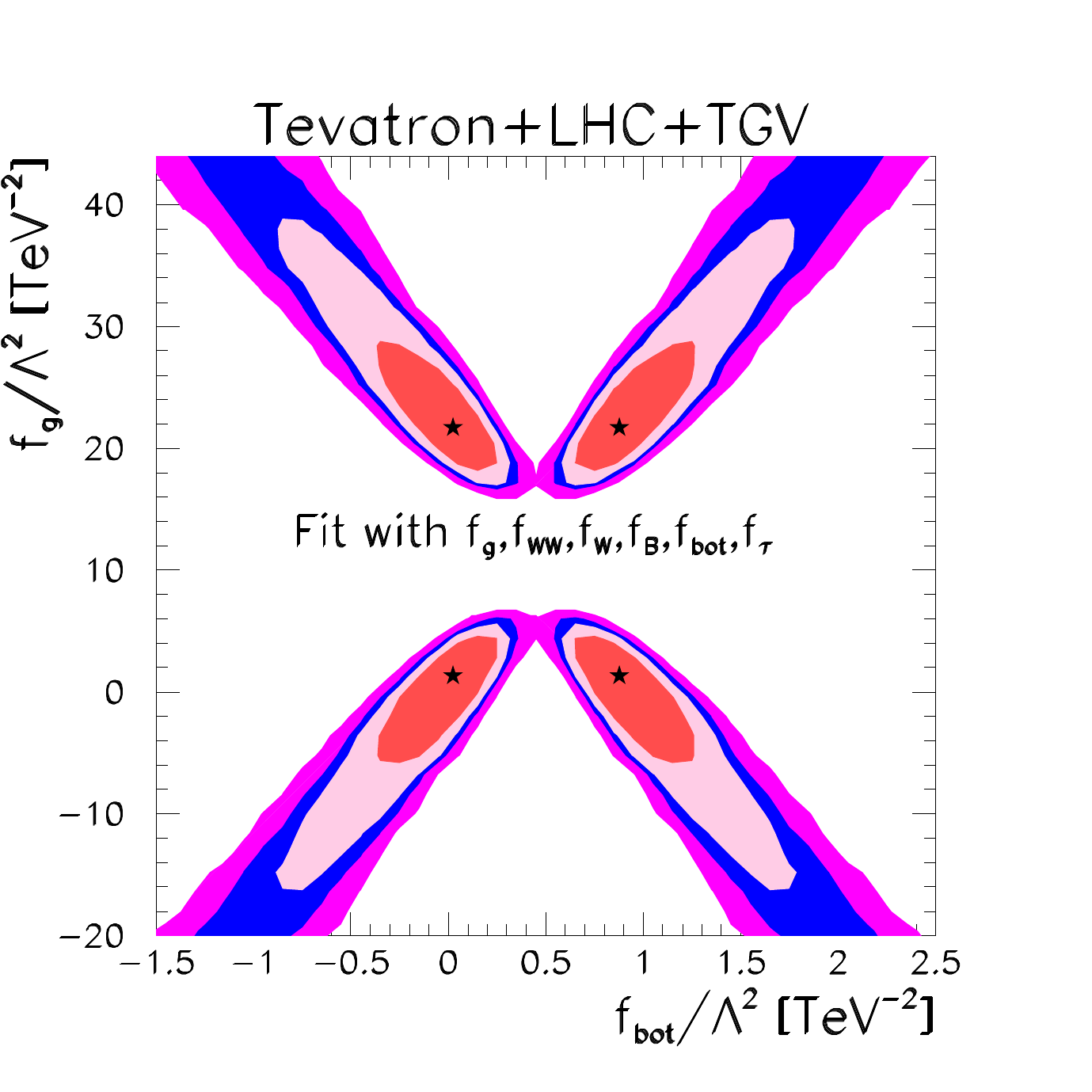}
 \caption{ We present the 68\%, 90\%, 95\%, and 99\% CL allowed
   regions for the  $f_{\rm bot} \times f_g$ plane when we fit the
   ATLAS, Tevatron, CMS, and TGC data while varying $f_g$, $f_{WW}$, $f_W$,
   $f_B$, $f_{\rm bot}$, and $f_\tau$.  The stars represent the global
   minima. Undisplayed parameters have been marginalized over.}
\label{fig:2a}
\end{figure}

Let us explore in more detail the two scenarios presented in
Figure~\ref{fig:2}, focusing on the effects of allowing for
the modification of Higgs couplings to fermions.  Comparing the
panels in the upper row with the ones in the lower row we can see the
allowed range of $f_g$ becomes much greater for the latter case, where
the range for $f_{\rm bot}$ is large as well.  This behavior has origins in
the fact that large $f_{\rm bot}$ causes the SMS branching ratio into pairs of
b--quark to approach 1, so in order to fit the data for any channel with a
final state F with $F\neq b\bar{b}$, the gluon fusion cross section
is enhanced in order to make up for the dilution of
$H\rightarrow F$ branching ratios. This occurs due to a strong
correlation between allowed values of $f_g \times f_{\rm
  bot}$~\cite{ourWork}.  This is illustrated in Fig.~\ref{fig:2a}
which depicts the strong correlation between the allowed values of
$f_g \times f_{\rm bot}$.

Still from Fig.~\ref{fig:2} we can see that allowing for $f_{\rm
  bot}\neq 0$ and $f_{\tau}\neq 0$ has a small impact on the
parameters affecting the SMS couplings to the electroweak gauge bosons
$f_W,$ $f_{WW}$  and $f_B$ as shown by comparing the corresponding
upper and lower panels. Concerning $f_\tau$, it does not possess any
strong correlation with other variables because the data on $pp \to h
\to \tau^+ \tau^-$ cuts off all strong correlations between $f_\tau$
and $f_g$. At the end, the introduction of $f_\tau$ has a little impact
on the determination of the other free parameters.

\begin{figure}
\centerline{\includegraphics[width=0.9\linewidth]{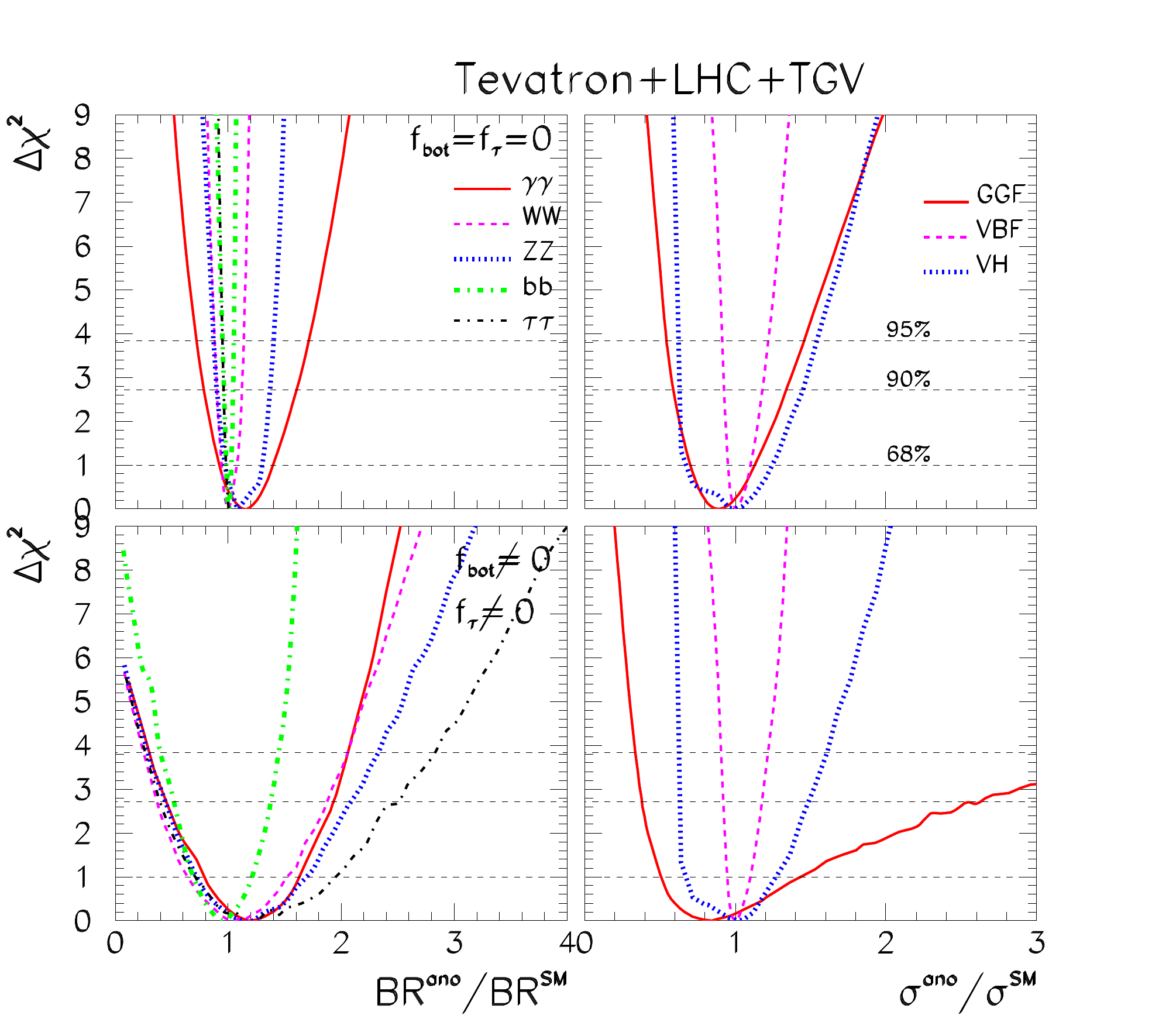}}
 \caption{ Chi--square as a function of the branching ratios (left
   panels) and the production cross sections (right panels) while
   considering all SMS collider(ATLAS, CMS and Tevatron) and TGC data.
   In the upper panels we have used $f_g$,  $f_W$, $f_{WW}$, and $f_B$ as
   fitting parameters setting $f_{\rm bot} = f_\tau =0$, while in the
   lower row we parametrize the data in terms of $f_g$, $f_{WW}$,
   $f_W$, $f_B$, $f_{\rm bot}$, and $f_\tau$. In the upper right
   panel, the dependence of $\Delta \chi^2$ on the branching ratio to
   the fermions not considered in the analysis arises from the effect
   of the other parameters in the total decay width.  }
\label{fig:3}
\end{figure}

\begin{figure}
\begin{minipage}{0.45\linewidth}
\centerline{\includegraphics[width=0.9\linewidth]{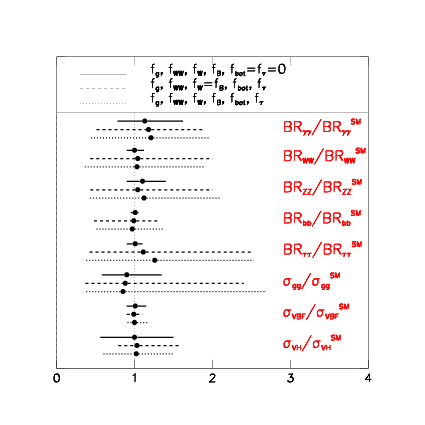}}
\end{minipage}
\hfill
\begin{minipage}{0.45\linewidth}
\centerline{\includegraphics[width=0.9\linewidth]{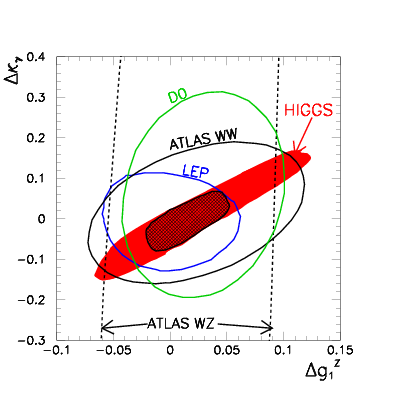}}
\end{minipage}
\caption{The left panel contains the 90\% CL limits on the SMS
  branching ratios and production cross sections. The right panel
  displays the 95\% CL allowed regions (2dof) on the plane
  $\Delta\kappa_\gamma \otimes \Delta g^Z_1$ from the analysis of the
  SMS data from LHC and Tevatron (filled region) presented in this
  work together with the relevant bounds from different TGC studies
  from collider experiments as labeled in the figure.  We also show
  the estimated constraints obtainable by combining these bounds
  (hatched region).  }
\label{fig:disc}
\end{figure}

We present in Fig.~\ref{fig:3} the chi-square dependence on branching
ratios and production cross sections for the two scenarios presented
in Fig.~\ref{fig:2}. The effect of $f_{\rm bot} \neq 0$ and $f_\tau
\ne 0$ on physical observables can be seen by comparing the
upper and lower rows of this figure: We can easily see that bounds
on branching ratios and cross sections get loosened, with the VBF and
VH production cross sections being the least affected quantities while
the gluon fusion cross section is the one which becomes less constrained.
The reason for this reduction of the constraints is the
strong correlation between $f_{\rm bot}$ and $f_g$ just mentioned. We
summarize the bounds on SMS production cross sections and branching
ratios in the left panel of Fig.~\ref{fig:disc} that shows that the SM
predictions for the SMS properties are in good agreement with the
available data.

The operators  ${\cal O}_W$ and ${\cal O}_B$ modify not only the SMS
interactions, but also give rise to TGC as we have already commented:
\begin{equation}
\Delta \kappa_\gamma = 
 \frac{g^2 v^2}{8\Lambda^2}
\Big(f_W + f_B\Big) 
\, ,\; 
\Delta g_1^Z= \frac{g^2 v^2}{8 c^2\Lambda^2}f_W 
\, ,\; 
\Delta \kappa_Z =   \frac{g^2 v^2}{8 c^2\Lambda^2}
  \Big(c^2 f_W - s^2 f_B\Big)\, ,
\label{eq:wwv}
\end{equation}
where $s$ $(c)$ stands for the sine (cosine) of the weak mixing angle.
Therefore, we use our framework to get bounds on TGC and we present
them in the right panel of Fig.~\ref{fig:disc}, where we can see that
the present SMS physics bounds on $\Delta\kappa_\gamma \otimes \Delta
g^Z_1$ show a non-negligible correlation.  This stems from the
correlation imposed on the high values of $f_W$ and $f_B$ from their
tree level contribution to $Z\gamma$ data, a correlation which is
transported to the $\Delta\kappa_\gamma \otimes \Delta g^Z_1$ plane.
Furthermore, the right panel of this figure also shows that the
present bounds on $\Delta\kappa_\gamma \otimes \Delta g^Z_1$ from the
analysis of SMS data are stronger than those coming from direct TGC
studies at the LHC.

The most important lesson that we can learn from the right panel of
Fig.~\ref{fig:disc} is the complementarity of the bounds on new
physics effects originating from the analysis of SMS signals and from
studies of the electroweak gauge--boson couplings~\cite{SMS-TGC}. To
assess the potential of this complementarity we combine the present
bounds derived from SMS data with those from the TGC analysis from
LEP, Tevatron and LHC shown in Fig.~\ref{fig:disc}. Clearly the
inclusion of the SMS data leads to stronger constraints on TGC. The
combined 1$\sigma$ 1dof allowed ranges are
\begin{eqnarray*}
-0.002\leq\Delta g^Z_1\leq 0.026,&
-0.034\leq\Delta\kappa_\gamma\leq0.034 & \\
{\rm which \; imply}&\;-0.002\leq\Delta\kappa_Z\leq0.029 &  \; .\nonumber  
\end{eqnarray*}

\section{Discussion and conclusions}

Here, we applied a bottom--up approach to describe possible departures of the SMS
couplings from the SM predictions.  Working in a model independent
framework the effects of the departures can be parametrized in terms of an effective
Lagrangian, more specifically we chose a basis of the dimension--six
operators such that we could use the largest possible dataset to
constrain the SMS couplings.  In this general framework the
modifications of the couplings of the SMS field to electroweak gauge
bosons are related to the anomalous triple gauge--boson
vertex~\cite{SMS-TGC}.  Our fit to the presently available data show
that the SMS branching ratios and cross sections are compatible with
the SM at 1$\sigma$ level. Moreover, the analysis of the Higgs boson
production data at LHC and Tevatron is able to furnish bounds on the
related TGC which, in some cases, are tighter than those obtained from
direct triple gauge--boson coupling analysis.  In the near future the
LHC collaborations will release their analysis of TGC with the larger
statistics of the 8 TeV run. The combination of those with the present
results from SMS data has the potential to furnish the strongest
constraints on new physics effects on the EWSB sector until further
luminosity can be accumulated \footnote{ As new data become available
  we will present the latest results in the site \url{
    http://hep.if.usp.br/Higgs }}.

\section*{Acknowledgments}

This work is supported by Conselho Nacional de Desenvolvimento
Cient\'{\i}fico e Tecnol\'ogico (CNPq), by Funda\c{c}\~ao de Amparo
\`a Pesquisa do Estado de S\~ao Paulo (FAPESP), by USA-NSF grant
PHY-09-6739, by CUR Generalitat de Catalunya grant 2009SGR502, by
MICINN FPA2010-20807 and consolider-ingenio 2010 program
CSD-2008-0037, by EU grant FP7 ITN INVISIBLES (Marie Curie Actions
PITN-GA-2011-289442), and by ME FPU grant AP2009-2546.


\end{document}